\documentclass[10pt,aps,prl,nofootinbib,twocolumn,showpacs]{revtex4}
\usepackage{amsmath,amsfonts}
\usepackage{graphicx,pstricks}
\newcommand{\arctanh}{\mathop{\rm arctanh}\nolimits}
\newcommand{\rot}{\mathop{\rm rot}\nolimits}
\newcommand{\f}{\frac}

\newcommand{\intl}{\int\limits}
\newcommand{\Div}{\mathop{\rm div}\nolimits}
\begin{document}
\title{Collective behavior in the system of
self propelling particles with nonholonomic constraints.}
\author{V.L. Kulinskii}
\email{koul@paco.net} \affiliation{Department for Theoretical
Physics, Odessa National University, 2 Dvoryanskaya St., 65026
Odessa, Ukraine}
\author{V.I. Ratushnaya, A.V. Zvelindovsky, D. Bedeaux}
 \affiliation{Colloid and Interface Science group, LIC, Leiden
University, P.O. Box 9502, 2300 RA Leiden, The Netherlands}
\begin{abstract}
We consider the dynamics of systems of self propelling particles
with nonholonomic constraints. A continuum model for a discrete
algorithm used in works by T. Vicsek et al. is proposed. For a
case of planar geometry the finite flocking behavior is obtained.
The circulation of the velocity field is found not to be conserved
as a consequence of the nonholomicity. The stability of ordered
motion with respect to noise is discussed. An analogy with the
kinetics of charges in superconductors is noted.
\end{abstract}
\pacs{05.65.+b, 47.32.-y, 87.10.+e} \maketitle
%%%%%%%%%%%%%%%%%%%%%%%%%%%%%%%%%%%%%%%%%%%%%%%%%%%%%%%%%%%%%%%%%%%%%%%%
The emergence of ordered structures in dynamic systems is a long
standing problem in physics. Generally speaking one deals with
dynamic phase transitions governed either by external or internal
noise. Most interesting is the arising of ordered motion caused by
the internal dynamics of the system. Recently, there has been a
growing interest in studying the collective behavior in systems of
self-propelling particles (SPP). One may distinguish systems of
two types:

The first one is for systems of ``unintelligent`` particles
interacting via real physical forces produced by the background in
which they are moving (micelles, bacteria, etc). Here the driving
forces are caused by the gradients of chemical factors
(concentrations, chem. potential etc.) or physical factors (light,
potential and dissipative fields etc.), which influence the
motion. In particular, the earth magnetic field is vital for the
orientation during long distance migration of biological species
like birds or turtles \cite{nat,sci} . It is clear that the
absence of conservation of translational and angular momentum is a
direct consequence of these external factors \cite{eb}.

The  second class is formed by systems of particles, which
interact via nonholonomic constraints imposed on their velocities.
We would like to stress that the nonholonomicity of such systems
clearly expresses the ``intelligent`` nature of the particles,
since such constraints need instant exchange of information
(visual or any other sensorial) between the particles and their
environment. This explains the coherent motion and arising ordered
patterns in the dynamics of systems like crowds, traffic or
flocking. Usual potential gradients or other physical forces are
not relevant for the collective behavior, though the particles
need some physical source of energy to sustain the constraints.
The physical origin of the nonholonomicity is the
force which acts on the particle due to its interaction
with the background (earth, air, liquid substrate etc.). Since we
are not interested in the dynamic degrees of freedom of the
background we lose this information. Such loss of dynamic
information leads to the breaking of the conservation of the
(angular) momentum and effectively to the nonholonomic
constraints.  In general it means that the system
with such constraints is not closed and therefore its dynamics is
not Hamiltonian, though it does not mean that the energy
dissipates. Rather a redistribution among the dynamic degrees of
freedom takes place. But the very form of the constraint is
determined by the ``intellect`` of a particle which uses the
information about the environment and moves accordingly. Note that
the numerical algorithm used in Ref.~\cite{ v2} modelled that kind
of systems. For shortness we call it the
Czir$\rm\acute{o}$k-Vicsek automaton or algorithm (CVA). There are
also modifications of this algorithm, which differ from the CVA by
inclusion of potential interparticle forces \cite{greg}, external
regular and stochastic fields \cite{scan}. We will consider CVA as
the minimal model for the collective behavior since the main cause
for the self-organization in the system is the nonholonomicity of
CVA dynamic rule.

To our knowledge none of the hydrodynamic models proposed earlier
is aimed to reflect the essence of the CVA - its conservative and
distinctly nonholonimic character. To a great extent such models
are modifications of the Navier-Stokes equation. Such an approach
is certainly valid for microorganisms floating in a medium. It is,
however, hardly adequate for the collective behavior of
"intelligent" boids, e.g. birds or drivers in traffic flow
\cite{traf}, when the individual behavior is determined rather by
the instant exchange of information with the environment than by
the action of some interparticle forces. The key point here is the
nonholonomicity of the system. In addition all the models based on
the modifications of the Navier-Stokes equation like
Ref.~\cite{tt,sr} include additional phenomenological terms which
generate the ``spontaneous`` transition to the state of ordered
motion. Such terms are added much in formal analogy with Landau's
theory of equilibrium continuous phase transitions and the notion
of an order parameter but any firm base for such an analogy is not
given. Thus the ordered state introduced in such a model is rather
an \textit{ad hoc} assumption than the natural consequence of the
underlying interparticle interactions. The interpretation of the
viscous term for such systems is also completely unclear.

In Ref.~\cite{tt} it was noted that the CVA can be considered as the
dynamic $XY$-model. It was concluded also that the cause of the
ordering is the convective term. The $XY$-model is a Hamiltonian
dynamic system. The latter property is the key point since for
Hamiltonian dynamic flow there is an ergodic measure. Due to
nonholonomic character the dynamic rule of the CVA even in the
static limit $|\mathbf{v}|\to 0$, i.e. when there is no any
transfer flow term like $\mathbf{v}\cdot\nabla$, has no canonical
Hamiltonian form. The general fact is that such systems do not
have an invariant measure with respect to the dynamic flow. In
fact the ordered state in the CVA appears as the fixed point
(attractor) of its averaging dynamic rule. Note that from the
point of view of the theory of dynamic systems the compactness of
the phase space (i.e. the space of positions and velocities) is
very important so the question about influence of the boundary
conditions usually used in simulations on the ordering the system
also needs discussion.

From such a nonholonomic point of view the appearance of ordered
motion in the CVA and similar dynamic systems is a trivial
consequence of their nonholonomicity. Indeed, the breaking of
conservation of the (angular) momentum is due to nonholonomic
constraints which as has been mentioned above mean that the system
is not closed.

%%%%%%%%%%%%%%%%%%%%%%%%%%%%%%%%%%%%%%%%%%%%%%%%%%%%%%%%%%%%%%%%%%%%%%%%

Here we discuss a hydrodynamic model which can be considered as
the continuous analogue of the discrete dynamic automaton proposed
in Ref.~\cite{v2} for the SPP system. It manifestly takes into
account all local conservation laws for the number of particles
and the kinetic energy. The self propelling force and the
frictional force are assumed to balance each other.

The algorithm used in Ref.~\cite{v2} corresponds to the following
equation of motion of a particle:
\begin{equation}\label{miceq}
\f{d}{dt}\mathbf{v}_i = \boldsymbol{\omega}_{i}\times \mathbf{v}_i,
\end{equation}
where $\boldsymbol{\omega}_i$ is the ''angular velocity'' of
$i$-th boid, which depends on what happens in the neighborhood. It
is assumed also that the number of particles is conserved.
Like algorithm in Ref.~\cite{v2}, Eq.~\eqref{miceq}
distinctly expresses the conservation of the kinetic energy.

From a physical point of view it is natural that the hydrodynamic
model corresponding to the CVA is based on the following
equations:
\begin{eqnarray}
  \f{d}{dt}\intl_{V} n(\mathbf{r},t)d V= & 0  \label{vic},\\
  \f{d}{d t} \intl_{V} n \mathbf{v}^2\,d V =&0  \label{kin}\,,
\end{eqnarray}
where $n(r,t)$ and $\textbf{v}(\mathbf{r},t)$ are the number
density and the Eulerian velocity. The volume $V$ moves along with
the velocity field. The first condition is the conservation of
number of particles. As usual we can rewrite this condition in the
differential form:
\begin{equation} \label{eqn}
\f{\partial n}{\partial t}+\Div\,\left(n\mathbf{v}\right)=0.
\end{equation}
The second constraint Eq.~\eqref{kin} means that the kinetic
energy of a Lagrange particle is conserved, i.e.
\begin{eqnarray}\label{kin1}
\f{d}{d t} \intl_{V} n \mathbf{v}^2\,d V&=&\intl_{V}
\mathbf{v}^2\left(\f{d\,n}{d\,t}+n \Div{\mathbf{v}}\right) dV
\nonumber\\
& & +\intl_{V} n\f{d\,\mathbf{v}^2}{dt} dV=0.
\end{eqnarray}
The first integrand vanishes due to the conservation of the
particle number. As a consequence the second integral also
vanishes for an arbitrary choice of the volume $V$ which in view
of the natural condition $n \ge 0$ leads to:
\begin{equation}\label{dv}
\f{d}{dt}|\mathbf{v}(\mathbf{r},t)|^2=0.
\end{equation}
This implies that a pseudovector field $\boldsymbol{\omega}$
exists such that
\begin{equation}\label{eqv}
\frac{d}{dt}\mathbf{v}\left(\mathbf{r},t\right)=
\boldsymbol{\omega}\left(\mathbf{r},t\right)
\times\mathbf{v}\left(\mathbf{r},t\right)\,.
\end{equation}
This equation, which is the continuous analogue of
Eq.~\eqref{miceq}, has now been derived from the conservation of
particle number and kinetic energy, Eq.~\eqref{vic} and
Eq.~\eqref{kin}.

We will model $\boldsymbol{\omega}$ by:
\begin{equation}\label{om0}
\boldsymbol{\omega}\left(\mathbf{r},t\right)=\int\,
K\left(\mathbf{r}-\mathbf{r'}\right)\,n(\mathbf{r'},t)\,\,
\rot\,\mathbf{v}(\mathbf{r'},t)\,d\mathbf{r'}\,,
\end{equation}
which has the proper pseudovector character and heuristically may
be considered as the continual analog of the CVA dynamic rule. There
are other possible choices like:
\begin{equation}\label{om1}
\boldsymbol{\omega}\left(\mathbf{r},t\right)=\int\,
\tilde{K}\left(\mathbf{r}-\mathbf{r'}\right)\nabla n\left(\mathbf{r'},t\right)
\times \mathbf{v}(\mathbf{r'},t)d\mathbf{r'}
\end{equation}
and combinations of the two. The averaging kernels $K$ and
$\tilde{K}$ should naturally decrease with the distance in
realistic models. We concentrate our discussion on the case
Eq.~\eqref{om0}. Equations \eqref{eqn} and \eqref{eqv} obviously
have the uniform ordered motion with a constant both density and
velocity as trivial solution.
%%%%%%%%%%%%%%%%%%%%%%%%%%%%%%%%%%%%%%%%%%%%%%%%%%%%%%%%%%%%%%%%%%%
One may consider Eq.~\eqref{eqv} as the equation of motion of a
charge in a magnetic field, where $\mathbf{v}$ is the charge
velocity and $\boldsymbol{\omega}$ is proportional to the magnetic
field. One may even include an ``electric field`` and
dissipative (collision) terms in Eq.~\eqref{eqv}:
\begin{equation}\label{eqv1}
\frac{d}{dt}\mathbf{v}\left(\mathbf{r},t\right)=\,\mathbf{f} +
\boldsymbol{\omega}\left(\mathbf{r},t\right)\times\mathbf{v}
\left(\mathbf{r},t\right) - \xi\mathbf{v}
\left(\mathbf{r},t\right)\,,
\end{equation}
where $\xi^{-1}$ is the mean free time. We will not
consider such an extended model here since algorithm
used in \cite{v2} is distinctly conservative. Further
we exploit the analogy of Eq.~\eqref{eqv} and the equation of
motion for charges in superconductors (see e.g. Ref.~\cite{dg}). In our
model a current density corresponding to the particle velocity at
some point $\mathbf{r}$ depends on ''magnetic field''
$\boldsymbol{\omega}$, i.e. ''vector potential'', at all
neighboring points $\mathbf{r'}$ within some region of coherence.
Therefore the relations \eqref{om0} or ~\eqref{om1} can be
considered as a corresponding nonlocal relations between the
current density and the vector potential in nonlocal Pippard's
electrodynamics of the superconductors. The situation in the
system under consideration is more complex. In electrodynamics of
superconductors the external magnetic field is the main cause of
the vortical motion of charges since their own
magnetic field is negligibly small and does not lead to formation
of the ordered motion. In our case the ''magnetic field''
$\boldsymbol{\omega}$ itself depends on the motion of ``charges``,
i.e. particles, and vise versa, which leads to a nonlinearity of
the system. Depending on the parameters one can
expect either the direct current state since the system is
conservative or vortical states like Meissner currents or
Abrikosov vortices.

Using such an analogy let us find the conditions for the existence
of the stationary vortical states.

We can rewrite Eq.~\eqref{eqv} in the following form:
\begin{equation}\label{om}
  \frac{\partial\, \boldsymbol{\omega}}{\partial\, t} +
  \frac{\partial\, \mathbf{W}}{\partial\, t}=
  \rot \left(\mathbf{v}\times\mathbf{W}\right)\,,
  \end{equation}
where $\mathbf{W}(\mathbf{r},t) = \rot\mathbf{v} -
  \boldsymbol{\omega}\,.$
Thus it follows that if $\mathbf{W}(\mathbf{r},t)$ is equal to $0$, then $
\partial\, \boldsymbol{\omega}/\partial\, t = 0$ and  therefore
$\boldsymbol{\omega}=\rot\mathbf{v}$ is independent of the time.
Such states are naturally interpreted as stationary translational
$\boldsymbol{\omega} = 0$ or rotational $\boldsymbol{\omega} \ne
0$ regimes of motion. For the model \eqref{om0} together with
$\mathbf{W}=0$ we get the integral equation:
\begin{equation}\label{om01}\int\,
K\left(\mathbf{r}-\mathbf{r'}\right)\,n(\mathbf{r'})\,\,
\rot\,\mathbf{v}(\mathbf{r'})\,d\mathbf{r'} =
\rot\,\mathbf{v}(\mathbf{r})\,,
\end{equation}
which determines such  states. Equation~\eqref{om01} gives stationary vortical motion, represented by the vector
field with $|\mathbf{v}|=const$. From here it follows
that the vorticity of the velocity field is an eigenstate of the
integral operator with $n(\mathbf{r})$ as the corresponding weight
factor. It should be noted that these stationary states do not
exhaust all stationary vortical states, since in general $\nabla
\mathbf{v}^2 \ne 0$.

%%%%%%%%%%%%%%%%%%%%%%%%%%%%%%%%%%%%%%%%%%%%%%%%%%%%%%%%%%%%%%%%%%%%%%%%
We further scale $K$ by multiplying with some $n^{\ast}$ and
similarly scale the density by dividing by $n^{\ast}$. Furthermore
we restrict our discussion to the simple case of a planar geometry
with averaging kernel in \eqref{om0} as $\delta$-functional:
\begin{equation}\label{k0}
  K(\mathbf{r} - \mathbf{r}') =s\,\delta(\mathbf{r} -
  \mathbf{r}')\,,\quad s = \pm 1.
\end{equation}
We will call this the local model. For this case
Eqs.~\eqref{eqn} and \eqref{eqv} take the form:
\begin{eqnarray}
\frac{\partial n}{\partial t}&+&
\Div\,\left(n\mathbf{v}\right)=0\label{nloc},\\
\frac{d}{dt}\mathbf{v}&=& s\,n\, \rot\mathbf{v}
\times\mathbf{v}\label{vloc}\,.
\end{eqnarray}
These equation \eqref{vloc} can be obtained as a special case of
the corresponding one in Ref.~\cite{tt}. In our work the
foundation of the terms is given. We do not use general symmetry
arguments to take all terms of a certain symmetry into account
irrespective of their physical meaning. The corresponding local
model with $\delta$-kernel for \eqref{om1} may be identified with
\textit{rotor chemotaxis} (i.e. caused by chemical field) force
introduced in Ref.~\cite{ac} if one takes into account simple
linear relation between field of food concentration and the number
density of boids, which is surely valid for low concentrations of
food and bacteria.

The parameter $s$ of the local model \eqref{vloc} distinguishes
different physical situations concerning the microscopic
nonholonomical constraint. To see this we find the stationary
radially symmetric solutions of \eqref{nloc} and \eqref{vloc}. As
usual, we search for the solutions of the form $n = n(r)$, $\mathbf{v} = v_{\varphi}(r)\mathbf{e}_{\varphi}$. The continuity equation \eqref{nloc} is satisfied trivially.
Substituting this into Eq.~\eqref{vloc} we finally obtain:
\begin{equation}\label{vort}
  v_{\varphi}(r) = \frac{C_{\rm st}}{2\pi r}
\exp\left(s\,\int\limits_{r_0}^{r}\,
\frac{1}{r\,n(r)}\,dr\right)\,,
\end{equation}
where $r_0$ is the cut-off radius of the vortex. This is the core
of the vortex. The constant $C_{\rm st}$ is determined by the circulation of the core $\oint\limits_{r=r_{0}} \mathbf{v}\,d\mathbf{l}  = C_{\rm st}\,.$

\begin{figure}
\begin{center}
\includegraphics[scale=0.5]{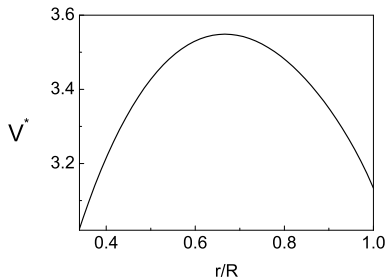}
\end{center}
\caption{Velocity $V^{*}(r/R)=2\pi R\,v_{\phi}(r)/C_{\rm st}$ in
the vortex of the local model with $n(r)$ given by Eq.~\eqref{n}
at $r_0/R=1/3$.}
\label{vr}
\end{figure}

The spatial character of the solution strongly depends on the
parameter $s$. If $s=-1$ the infinitely extended distributions for
$n(r)$ are allowed, e.g. $n(r)\propto r^{-\alpha}\,,
\,\,\alpha>0$. They lead to localized vortices with exponential
decay of angular velocity. If $s=+1$ only compact distributions,
i.e. $n(r)\equiv 0$ outside some compact region, are consistent
with the finiteness of the total kinetic energy i.e. they
corresponds to finite number of particles $\int n\, dV < \infty\,$.
As an example we may give:
\begin{equation}\label{n}
n(r) = \left\{\begin{array}{ccc}
  \sqrt{\f{r_0}{R-r}}\,,& \,\,r_0< r<R, \\
  0\,,&  \text{otherwise}&\,.
\end{array}\right.
\end{equation}
Substituting Eq.~\eqref{n} into Eq.~\eqref{vort} one obtains:
\begin{equation}
v_{\varphi}=\f{C_{\rm st}}{2\pi
r}\exp\left[2\sqrt{\f{R}{r_0}}\left(\sqrt{1-\f{r}{R}}-
\arctanh\sqrt{1-\f{r}{R}}\right)\Bigg|_{r_{0}}^{r}\right].
\end{equation}
The corresponding component of the velocity $ \mathbf{v} =
v_{\phi}\,\mathbf{e}_{\phi}$ for such a case is shown on
Fig.~\ref{vr}.
%%%%%%%%%%%%%%%%%%%%%%%%%%%%%%%%%%%%%%%%%%%%%%%%%%%%%%%%%%%%%%%%%%%%%%%%

Taking the rotation on both sides of Eq.~\eqref{vloc} one obtains
the following equation for the vorticity in a case of planar
geometry:
\begin{equation}\label{rott}
\rot\f{d\mathbf{v}}{dt}=-s\left(\frac{\partial\, n}{\partial\,
t}\rot{\mathbf{v}}
 - n\, (\mathbf{v}\cdot\nabla)
\rot{\mathbf{v}}
 \right).
\end{equation}
This implies that for $s=+1$ the vorticity is damped by
compression along the flow and therefore such a flow is stable
with respect to the vortical perturbations. For $s=-1$ the
vorticity is damped by expansion.

The hydrodynamics of nonholonomic fluid under consideration
differs essentially from the potential dynamics of ideal fluids
\cite{ll} due to the nonpotential character of the equation of
motion \eqref{eqv}. In ideal fluids the nontrivial stationary
flows are only possible under the influence of an external force
(pressure gradient or reaction of boundaries). Moreover the
vorticity in the nonholonomic fluid is not coserved. This is not a
case for usual ideal fluids where the vorticity is the integral of
motion.

The first term in right-hand side of Eq.~\eqref{rott} shows the
influence of compression on the evolution of the vorticity. The
second term represent the modified spatial transfer of the
vorticity along the flow. Since this term can be excluded locally
by the choice of the instant local reference frame which moves
accordingly with the flow we consider the first term as the main
source of the vorticity. In such an approximation we can write:
\begin{equation}\label{rottt}
\rot\f{d\mathbf{v}}{dt}=-s\frac{\partial\, n}{\partial\,
t}\rot{\mathbf{v}}.
\end{equation}
The circulation is defined by
\begin{equation}\label{ct}
C(t)=\oint\mathbf{v}d\mathbf{l}=\intl_{S}\rot\mathbf{v}d\mathbf{S},
\end{equation}
where the integration contour and the corresponding surface area
move along with the velocity field. The time derivative of the
circulation is:
\begin{equation}\label{rot2}
\f{d}{dt}\,C =\intl_{S}\rot\f{d\mathbf{v}}{dt}d\mathbf{S}\,.
\end{equation}
Thus the circulation does not conserve in contrast to the ideal
fluid model, where the microscopic interactions are of holonomic
character.

%%%%%%%%%%%%%%%%%%%%%%%%%%%%%%%%%%%%%%%%%%%%%%%%%%%%%%%%
The total momentum $\mathbf{P} =\int n\,\mathbf{v}\,dV$ does not
conserve too. For the local model we can write:
\begin{equation}\label{pt}
  \f{d}{d t}\mathbf{P} =\int
 s\, n^2 \rot{\mathbf{v}}\times\mathbf{v}\,dV\,.
\end{equation}
From Eq.~\eqref{pt} it follows that the damping of vortical part
of the velocity leads to the formation of the state of uniform
motion with $\mathbf{P}=\rm const$.

%%%%%%%%%%%%%%%%%%%%%%%%%%%%%%%%%%%%%%%%%%%%%%%%%%%%%%%%%%%%%%%%%%%%%%%%
Here we consider the influence of noise on the
stability of the flow with respect to the vortical perturbation.
It is clear that instability with respect to such perturbations
drives the system to disordered state. Inclusion of stochastic
noise can be done in a way analogous to that used in
Ref.~\cite{v2}: $\boldsymbol{\omega}\left(\mathbf{r},t\right)=
  \boldsymbol{\omega}_0\left(\mathbf{r},t\right)+
  \delta\boldsymbol{\omega}\left(\mathbf{r},t\right)\,,$
where $\boldsymbol{\omega}_0=sn\rot\mathbf{v}$ is the same
contribution as before and $\delta\boldsymbol{\omega}$ is the
stochastic contribution. These fluctuations lead to fluctuations
of the density and velocity fields. Replacing $\partial n/\partial
t$ by an average value $1/\tau$ plus a fluctuating contribution
$\delta L(t)$ in Eq.~\eqref{rottt} one
obtains for the above described local model with $s=+1$:
\begin{equation}\label{v1}
\frac{d\, }{d\, t}  C = -\left(\frac{1}{\tau}+\delta L\right) C\,,
\end{equation}
The simplest model for the noise is the Gaussian white noise approximation:
\begin{equation}
\left\langle\, \delta L(t)\delta L(t') \,\right\rangle =
2\,\Gamma\,\delta(t-t')\,.
\end{equation}
The stochastic equation
\eqref{v1} has the solution:
\begin{equation}\label{solv1}
C(t) =
C_0\exp\left(-\frac{t}{\tau}\right)\,
\exp\left(-\mathcal{W}(t)\right)\,,
\end{equation}
where $\mathcal{W}(t)=\int\limits_{0}^{t}\delta
L(t')\,dt'$ \,is the Wiener process \cite{vank}. Averaging over the
realization of the stochastic process we get the averaged
evolution of the vorticity:
\begin{equation}\label{solv2}
\left\langle\, C(t) \,\right\rangle
 =
C_0\exp\left(-\,\frac{t}{\tilde{\tau}}\right)\,,\quad \tilde{\tau}
= \frac{\tau}{1-\tau\Gamma},
\end{equation}
where $\tilde\tau$ is the relaxation time of the circulation in
the system. For large enough noise $\tau\,\Gamma > 1$ the system
becomes unstable. For $\tau\,\Gamma \le 1$ the system is stable
and the circulation decays to zero. When $\tau\Gamma$ increases to
$1$ the relaxation time $\tilde\tau$ goes to infinity, a result
similar to critical slowing down near the critical point.  These
estimates are modified when other terms, which have been
neglected, are taken into account but we believe that
qualitatively the obtained results remain unchanged.

It should be noted that the average kernels can also contain noise
contributions. In view of the above results for the character of
stationary states and vorticity relaxation, which depends on the
sign of parameter $s$, this case needs a more thorough
investigation.

In conclusion we have constructed a continuum SPP model with
particle number and kinetic energy conservation. We found in 2D
that vortical solutions exist for the model and that they show a
finite flocking behavior. These solutions, which qualitatively
reproduce some results of Ref.~\cite{v2}, were obtained without
imposing any boundary conditions on the velocity field.
The nonholonomic constrains were found to lead to a circulation which
was not conserved. The influence of noise on the stability of the
system was discussed.
%%%%%%%%%%%%%%%%%%%%%%%%%%%%%%%%%%%%%%%%%%%%%%%%%%%%%%%%%%%%%%%%%%
\section{acknowledgements}
Vladimir Kulinskii thanks NWO (Nederlandse Or\-ga\-ni\-sa\-tie
voor Wetenschappelijk Onderzoek) for a grant, which enabled him to
visit Dick Bedeaux's group at the Leiden University.

\end{document}